\documentclass{kapproc} 

  \usepackage[T1]{fontenc}

\usepackage{procps} 

\usepackage[dvips]{graphicx}

\upperandlowercase

\kluwerbib 

\begin{document}



\articletitle[The FIR-Radio Correlation using {\it Spitzer}]
{van der Kruit to Spitzer: A New Look at the FIR-Radio Correlation}

\chaptitlerunninghead{The FIR-Radio Correlation using {\it
    Spitzer}\hfill E. J. Murphy et al.}













--------------

\author{E.J. Murphy\altaffilmark{1}, R. Braun\altaffilmark{2}, 
         G. Helou,\altaffilmark{3}, L. Armus,\altaffilmark{3},
         J.D.P. Kenney\altaffilmark{1}, and the SINGS team}
\altaffiltext{1}{Department of Astronomy, Yale University,
  New Haven, CT 06520; murphy@astro.yale.edu}
\altaffiltext{2}{ASTRON, P.O. Box 2, 7990 AA Dwingeloo,
  The Netherlands}
\altaffiltext{3}{California Institute of Technology, MC
  314-6, Pasadena, CA 91101}



%

\begin{abstract}
We present an initial look at the far infrared-radio correlation
within the star-forming disks of four nearby, nearly face-on galaxies
(NGC~2403, NGC~3031, NGC~5194, and NGC~6946).
Using {\it Spitzer} MIPS imaging and WSRT radio continuum data, we are able
to probe variations in the logarithmic 70~$\mu$m/22~cm ($q_{70}$) flux
density ratios across each disk at sub-kpc scales.  
We find general trends of decreasing $q_{70}$ with declining surface
brightness and with increasing radius. 
We also find that the dispersion in $q_{70}$ within galaxies is
comparable to what is measured {\it globally} among galaxies at around
0.2 dex.  
We have also performed preliminary phenomenological modeling of cosmic
ray electron (CR$e^{-}$) diffusion using an image-smearing technique, 
and find that smoothing the infrared maps improves their correlation 
with the radio maps.  
The best fit smoothing kernels for the two less active star-forming
galaxies (NGC~2403 and NGC~3031) have much larger scale-lengths than
that of   the more active star-forming galaxies (NGC~5194 and
NGC~6946).
This difference may be due to the relative deficit of recent CR$e^{-}$
injection into the interstellar medium (ISM) for the galaxies having
largely quiescent disks.
A more complete discussion of this proceedings article can be found in
\cite{ejm05}.
\end{abstract}

\section{Introduction}
Soon after the Westerbork Radio Synthesis Telescope (WSRT) went
online, \cite{vk71} made the first report of a correlation between
1415~MHz radio and 10~$\mu$m mid-infrared luminosities for a
sample of Seyfert galaxy nuclei.  
This discovery was soon extended to the nuclei of normal spirals,
though the dispersion was found to be smaller in the correlation for
the original sample of Seyferts (\cite{vk73}). 
It was not until the coming of IRAS that the optically thin
radio continuum emission from galaxies was found to be better
correlated with the far infrared (FIR) dust emission of galaxies
{\it without} an active galactic nucleus (AGN) (\cite{de85,gxh85}).  

The connection between radio and infrared emission from galaxies is
that they are both powered by massive stars, as pointed out originally
for starbursts by \cite{har75}.
Young massive stars, which heat up dust to provide the FIR emission,
are the same stars which end as supernovae (SNe) and bring about
the synchrotron emission observed at radio wavelengths. 
If this general picture is correct, the fact that the mean free path
of UV photons ($\sim$100~pc) which heat the dust is much less than the
diffusion length for a CR$e^{-}$ ($\sim$1-2~kpc) suggests that the
radio image should resemble a smeared version of the infrared image. 
This idea was first introduced by \cite{bh90}, who attempted
to model the propagation of CR$e^{-}$s by smearing IRAS data of
galaxies using parameterized kernels containing the physics of the
CR$e^{-}$ propagation and diffusion, to better match the morphology of
the corresponding radio data. 
 
In an attempt to better understand the FIR-radio correlation, we 
are using MIPS infrared data from the {\it Spitzer} Infrared Nearby
Galaxies Survey (SINGS) legacy science project (\cite{rk03}).  
In this initial study, we examine the variations of the FIR-radio
correlation on sub-kpc scales within four of the nearest face-on
galaxies in the SINGS sample: NGC~2403, NGC~3031 (M81), NGC~5194
(M51a), and NGC~6946.  
Using high resolution {\it Spitzer} imaging, we are also able to test
the smearing model of \cite{bh90} with greater accuracy, at higher
spatial resolution, and in more galaxies, with the aim of gaining
better insight into CR$e^{-}$ diffusion and confinement within galaxy
disks.    
For a complete discussion, we direct the reader to \cite{ejm05}.

\section{Results}
\subsection{$q_{70}$ Maps}
\begin{figure}[!ht]
\center{
\scalebox{.31}{
\includegraphics{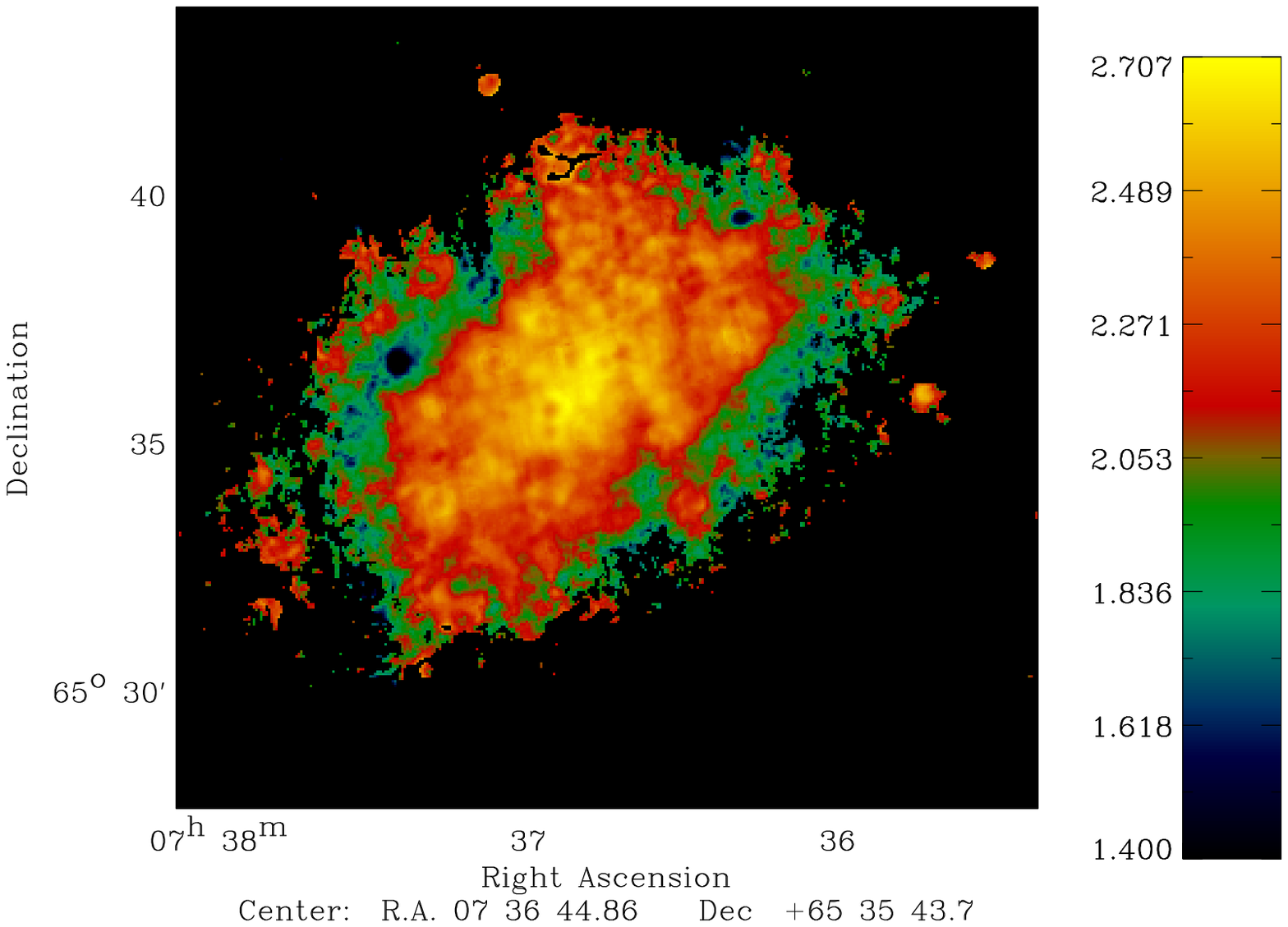}
\includegraphics{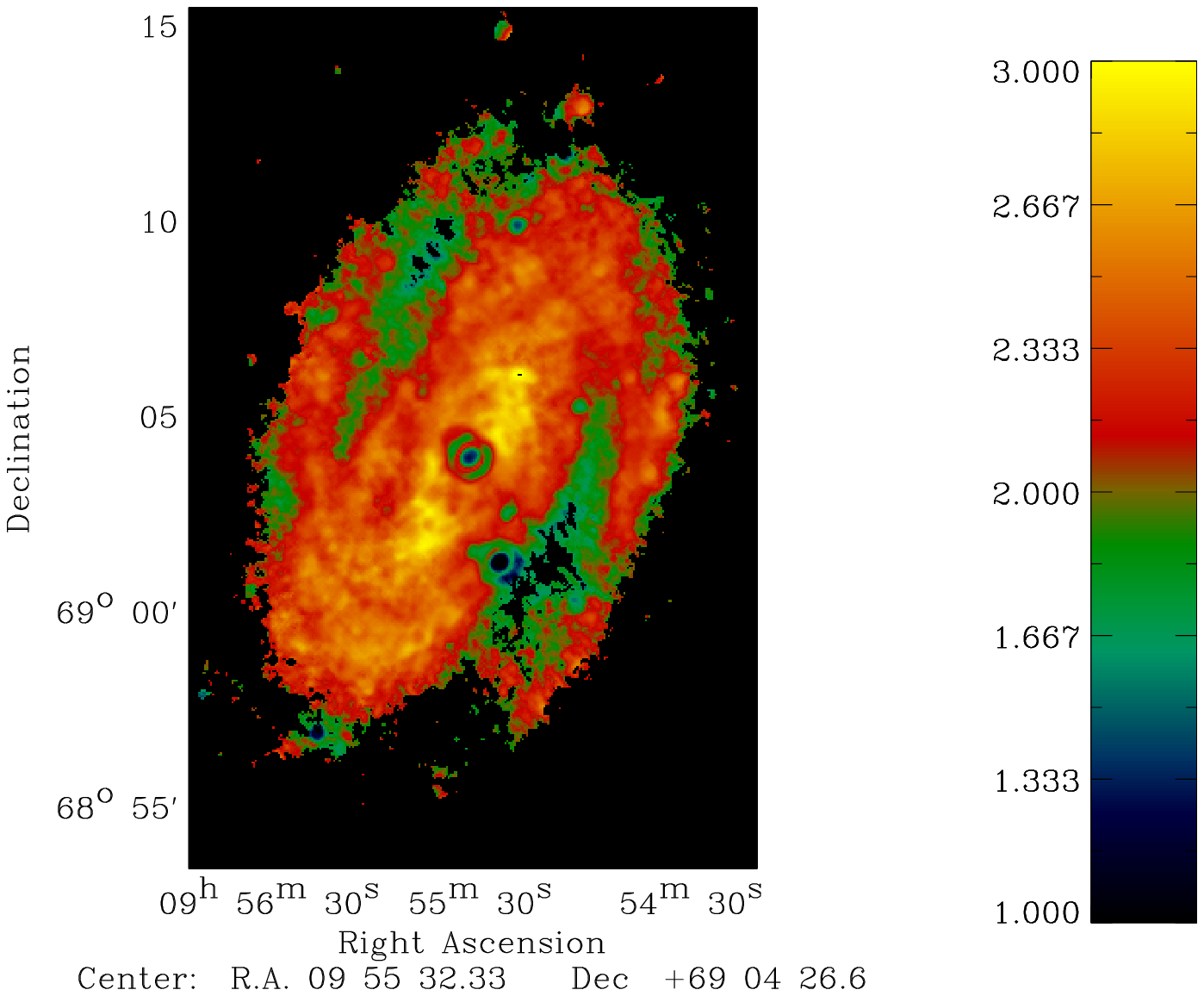}}\\
\scalebox{.31}{
\includegraphics{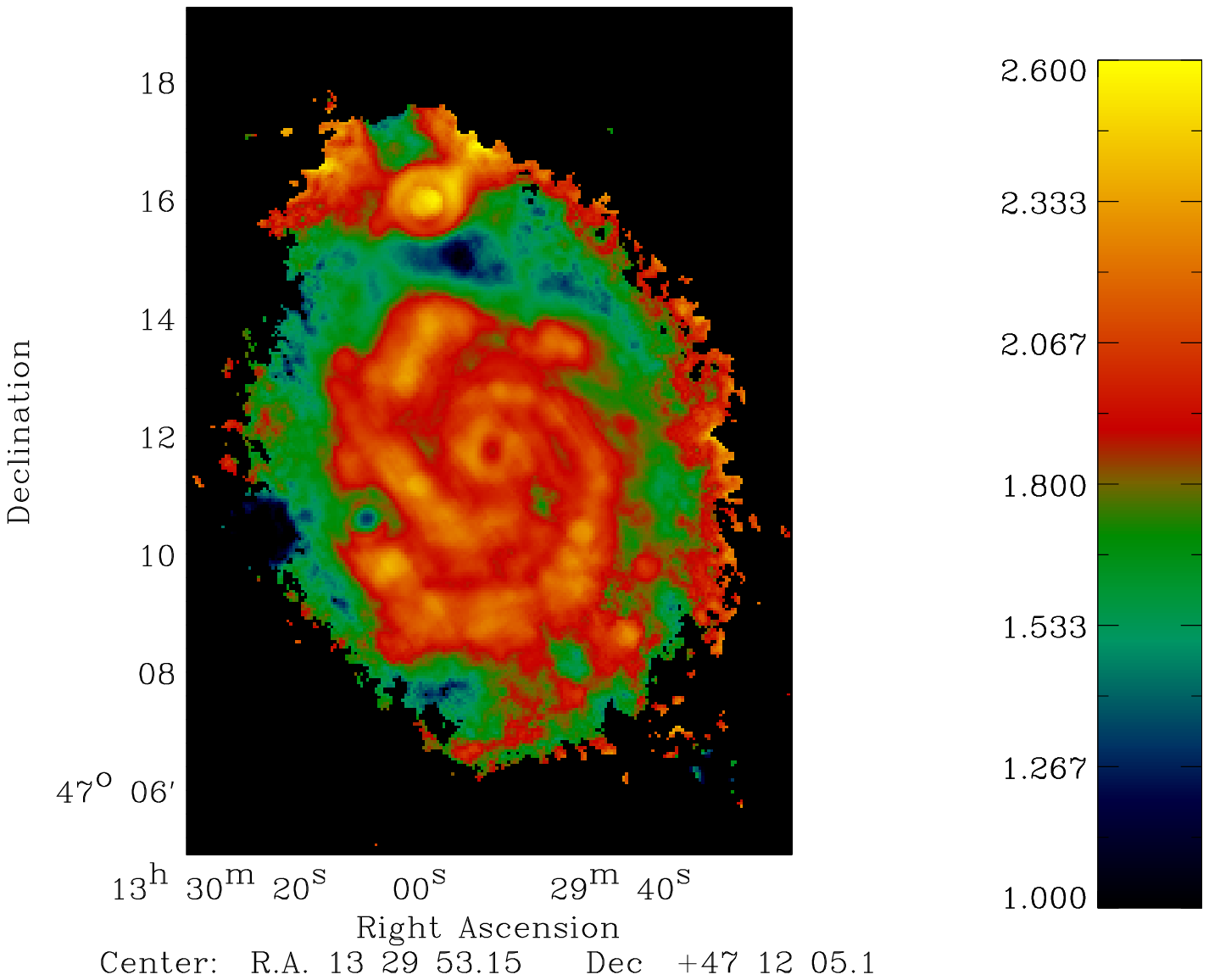}
\includegraphics{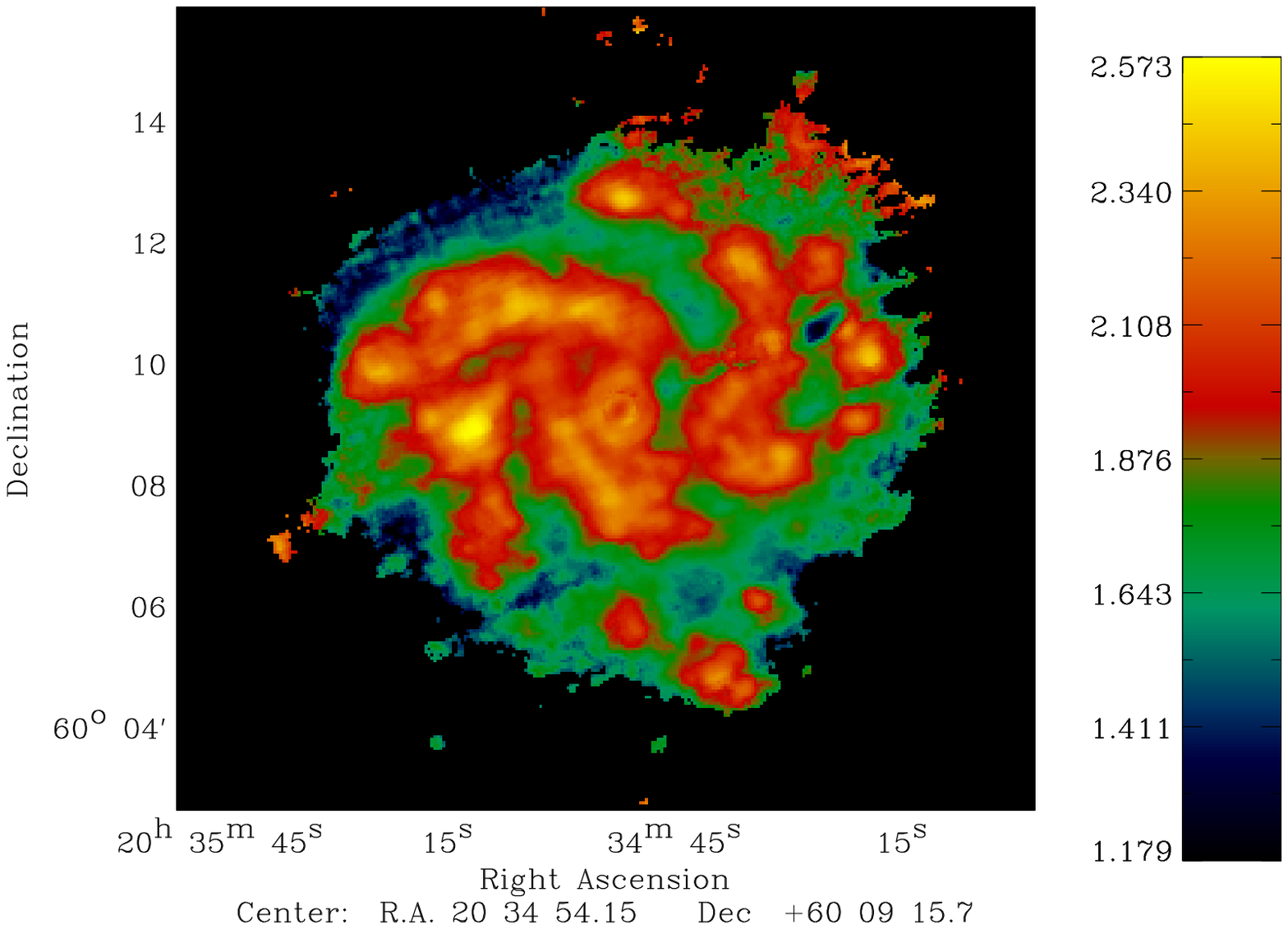}}
} 
\caption{\protect\inx{\label{qmaps}} $q_{70}$ maps for pixels having
  a 3-$\sigma$ detections in both the input radio and 70~$\mu$m maps.
  The $q_{70}$ ratios are enhanced along spiral arms and on H II
  regions due to 70$\mu$m emission being more strongly peaked than the
  22~cm emission.}
\end{figure}

In Figure \ref{qmaps} we plot \(q_{70} \equiv
\log\left(\frac{f_{\nu}(70~\mu\rm{m})[\rm{Jy}]}
	 {f_{\nu}(22~\rm{cm})[\rm{Jy}]}\right)\) for each galaxy.  
We find that elevated $q_{70}$ ratios are associated with bright
structures in the infrared and radio images of each galaxy.  
The most obvious cases are seen for the bright spiral arms of NGC~3031,
NGC~5194 and NGC~6946.
All three galaxies show enhanced $q_{70}$ ratios along their arms,
with local peaks centered on H II regions, and depressed ratios
located in the quiescent inter-arm and outer-disk regions of each
galaxy.   
For NGC~2403, which does not have a grand-design spiral morphology, we
still see $q_{70}$ peaks associated with H II regions.

To quantify radial variations in $q_{70}$, along with any dependencies
of $q_{70}$ on the 70~$\mu$m surface brightness within each galaxy, we
perform 
aperture photometry using apertures having diameters equal to the FWHM
of the MIPS 70~$\mu$m PSF (i.e. 17'').
The measured dispersion of $q_{70}$ within each galaxy is less than
$\sim$0.25~dex.
We find a general trend of increasing $q_{70}$ ratios with increasing
70~$\mu$m surface brightness and decreasing radius. 
The residual dispersion around the trend of $q_{70}$ with increasing
radius is found to be larger than the residual dispersion around the
trend of $q_{70}$ with increasing 70~$\mu$m surface brightness by
$\sim$0.1~dex.  
This suggests that the star formation sites within the disk are more
important in determining the overall appearance of the $q_{70}$
maps compared to the exponential profiles of the disks themselves.

\subsection{Infrared/Radio Relations Inside and Among Galaxies}
\begin{figure}[!ht]
\center{
\scalebox{.5}{
\includegraphics{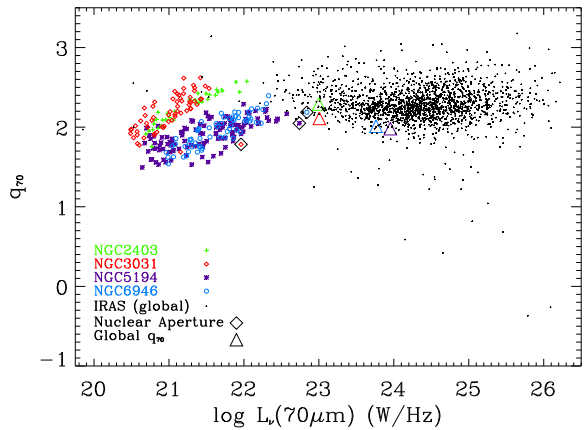}
}} 
\caption{\protect\inx{\label{globloc}}  Matched 1.5 kpc aperture
  measured $q_{70}$ ratios for each sample galaxy plotted with global
  $q_{70}$ ratios estimated from the data presented by \cite{yun01}.
  While the global $q_{70}$ ratios are constant with 70~$\mu$m
  luminosity, the local sub-kpc $q_{70}$ ratios are not.}
\end{figure}

In Figure \ref{globloc} we show our local sub-kpc $q_{70}$ ratios
together with global $q_{70}$ values for 1752 galaxies from the study
of \cite{yun01}.  
Using their cataloged 1.4~GHz NRAO VLA Sky Survey (NVSS) and 60 and
100~$\mu$m IRAS data, we converted IRAS based $q_{60}$ values to {\it
  Spitzer} $q_{70}$ values using the SED models of \cite{dd02}.  
We find that the dispersion in the local sub-kpc $q_{70}$ ratios
within our sample galaxies is comparable to that of global ratios.  
In the \cite{yun01} sample, $q_{70}$ appears to be roughly constant
with increasing galaxy luminosity while, within each disk, $q_{70}$
ratios clearly increase with the 70~$\mu$m luminosity.  
This difference in the observed trend between $q_{70}$ and luminosity
within and among galaxies can be explained by the diffusion of
CR$e^{-}$ within the galaxy disks.  

\section{Image-Smearing and Cosmic Ray Diffusion}
To help understand the underlying physics of the correlation, we look
to see whether the image-smearing model works to improve the
correlation between the infrared and radio morphologies of galaxies.  
Determining the functional form of the best fit smearing kernel
provides insight into the propagation and diffusion characteristics
of cosmic rays within galaxy disks.
In this study we examine the behavior for Gaussian and exponential
kernels oriented in the planes of the sky and each galaxy disk.

\begin{figure}
\center{
\scalebox{.34}{
\includegraphics{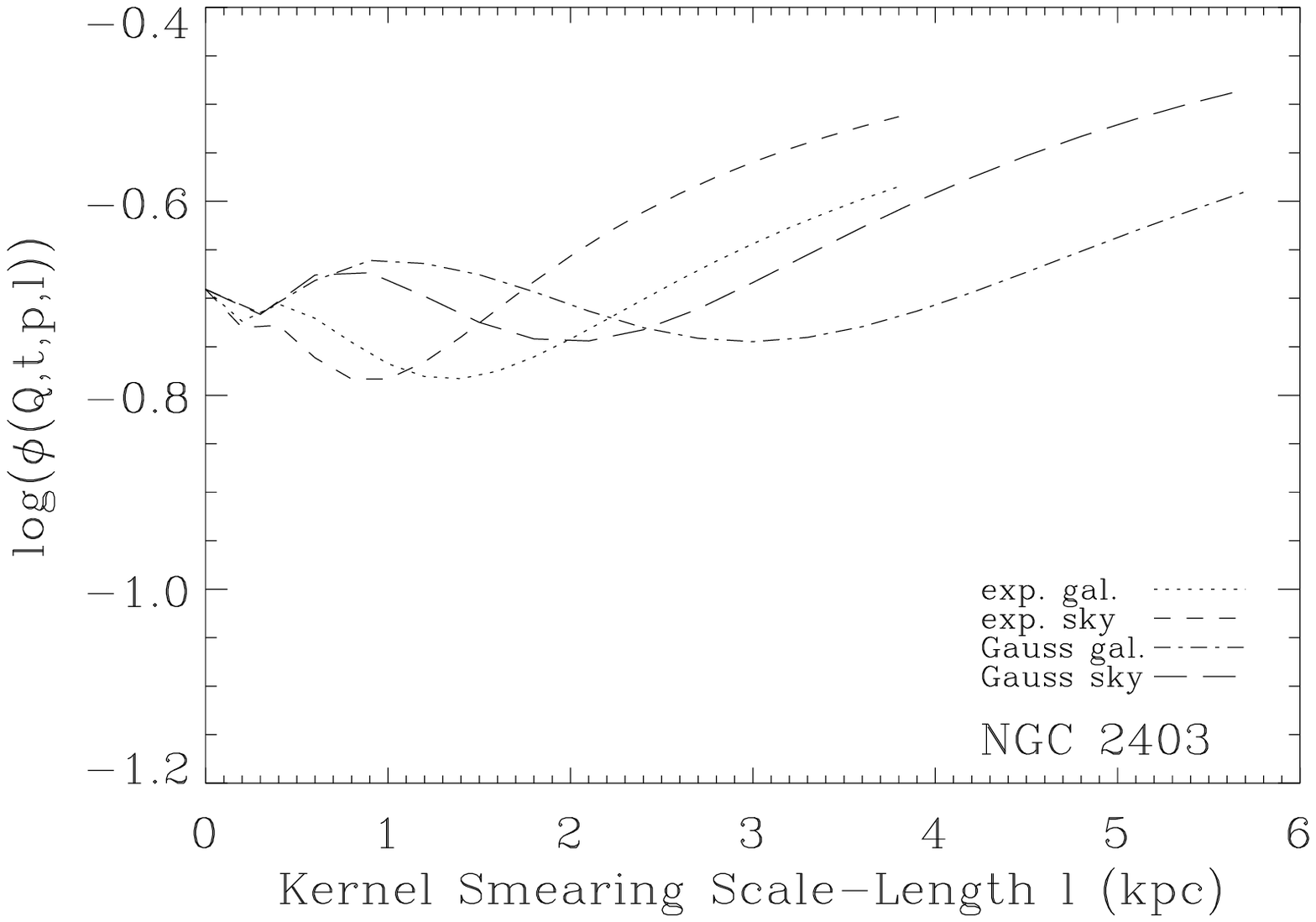}
\includegraphics{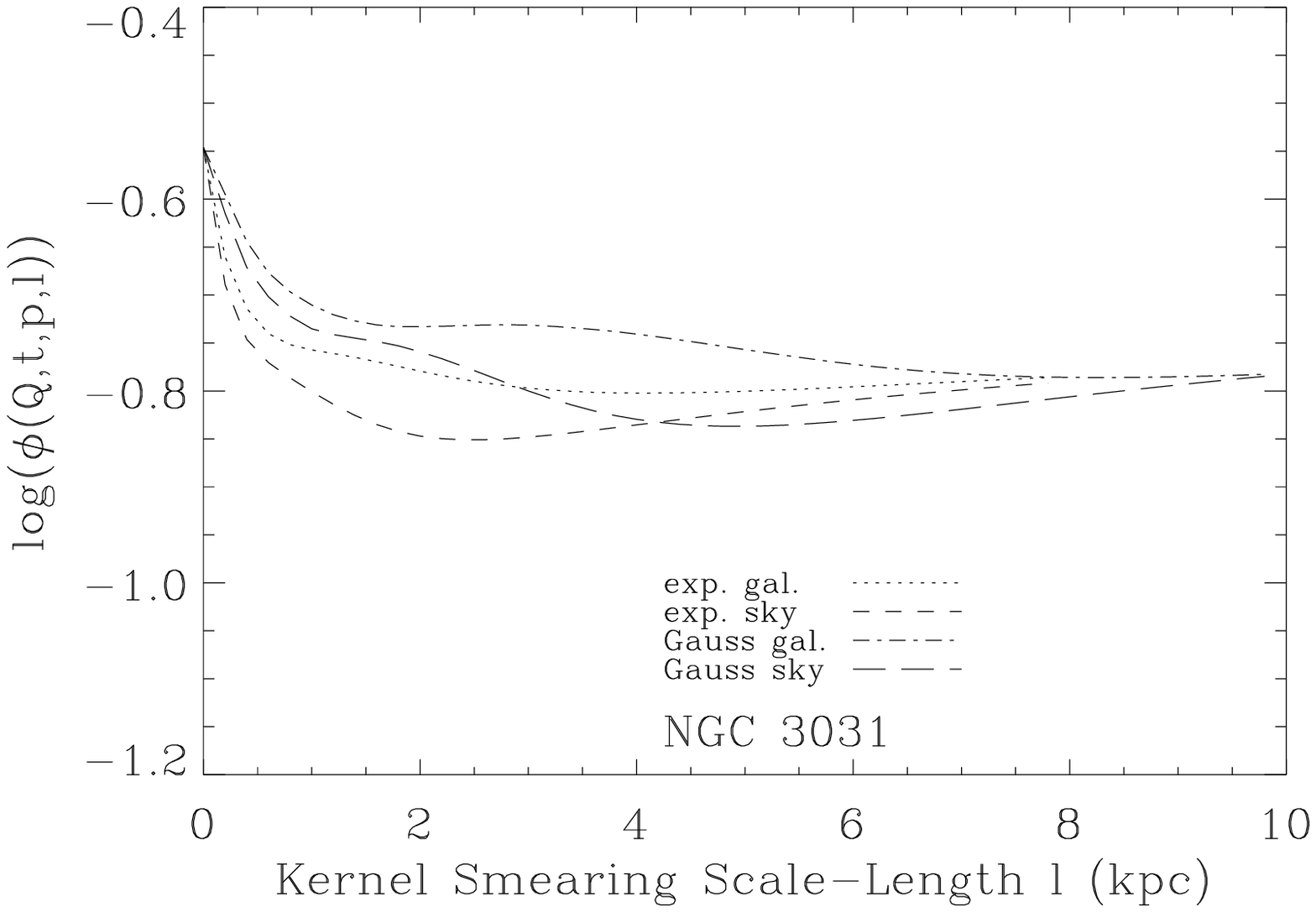}}\\
\scalebox{.34}{
\includegraphics{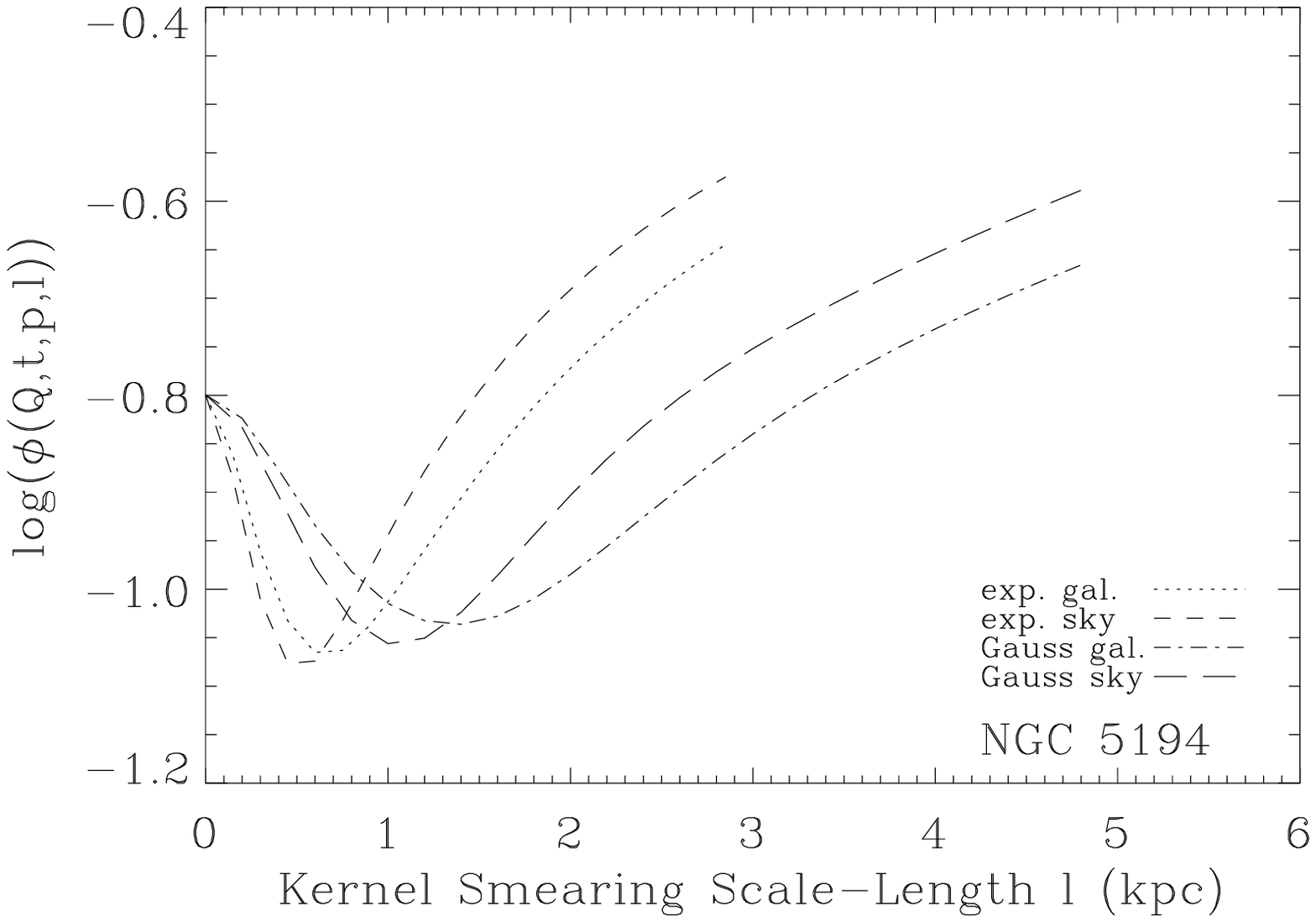}
\includegraphics{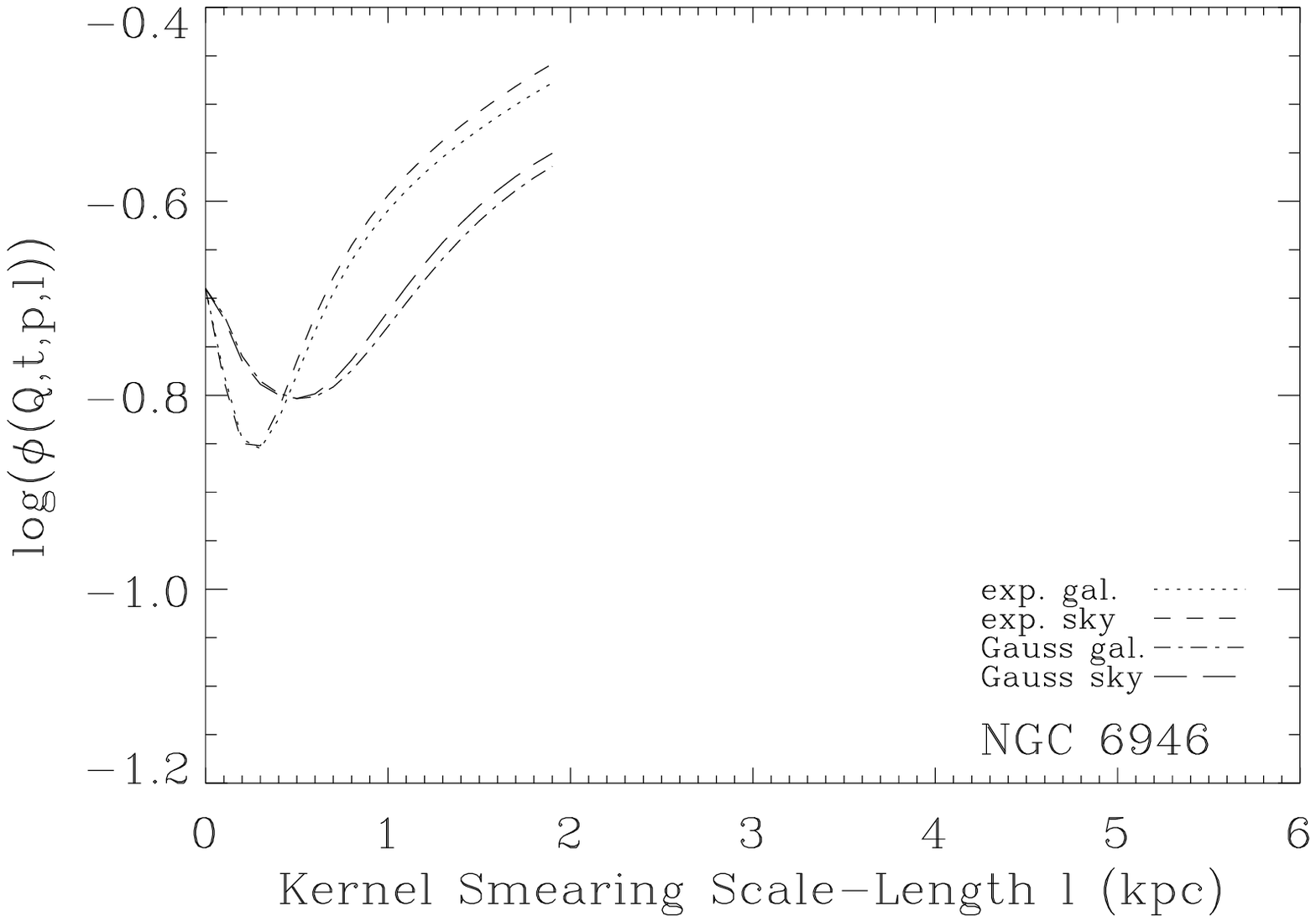}}
} 
\caption{\protect\inx{\label{res70}} Residuals of observed radio maps
    with smeared 70~$\mu$m images, defined by $\phi(Q,t,p,l)$, as
    a function of smearing scale-length shown for each kernel per
    galaxy.  Residuals for the active star-forming galaxies are
    minimized by kernels having $<$1~kpc scale-lengths while the less
    active star-forming galaxies are better fit by those with
    scale-lengths $>$1~kpc.} 
\end{figure}

We define the residuals between the smeared 70~$\mu$m map (
$\tilde{I}({t,p,l})$:$t=$shape, $p=$orientation, and $l$=scale-length)
and the observed 22~cm map (R) by,
\begin{equation}
\footnotesize
\phi(Q,t,p,l) = \frac{\sum[Q^{-1}\tilde{I}_{j}(t,p,l) -
    R_{j}]^{2}}{\sum R_{j}^2},
\label{phi}
\end{equation} 
where \(Q=\frac{\sum \tilde{I}_{j}({\bf r})}{\sum R_{j}({\bf r})}\)
and $j$ indexes each pixel.  
In Figure \ref{res70} we plot $\log(\phi)$, as a function of the
scale-length ($l$), for each galaxy and find that the image-smearing
technique improves the overall correlation  between the radio map and
the 70~$\mu$m image by an average of $\sim$0.2~dex.  
Unlike \cite{mh98}, we find exponential kernels are preferred
independent of projection over Gaussian kernels, suggesting that
additional processes such as escape and decay appear necessary to
describe their evolution through the galaxy disks.

We find distinct behavior between our active and less active
star-forming galaxies.   
Figure \ref{res70} shows that scale-lengths $<$1~kpc work best to
improve the correlation for the more active star-forming galaxies
(NGC~5194 and NGC~6946: SFR$\geq$4 M$_{\odot}$/yr), while
scale-lengths $>$1~kpc work best for the less active star-forming
galaxies (NGC~2403 and NGC~3031: SFR $<$1M$_{\odot}$/yr). 
We also find that by smearing the 70~$\mu$m images, we are able to
reduce the slope of the observed non-linearity in $q_{70}$ with
70~$\mu$m surface brightness by $>$25\% for each galaxy.
This suggests that the non-linearity may be due to the diffusion of
CR$e^{-}$s and is not expected to be found in the {\it global}
correlation when integrating the flux over entire galaxies.

\section{Conclusion}
In examining the behavior of $q_{70}$ within the disks of our sample
galaxies, along with our image-smearing analysis, we find: 
\begin{enumerate}
\item 
  An empirical trend of $q_{70}$ decreasing with declining 70~$\mu$m
  surface brightness and increasing radius to be a general 
  property within the galaxies.
  However, the dispersion measured in $q_{70}$ at constant 70~$\mu$m
  surface brightness is found to be smaller than at constant radius by
  $\sim$0.1 dex, suggesting that star formation sites are more
  important in determining the $q_{70}$ disk appearance than the
  underlying exponential disks.
\item
  The dispersion in the {\it global} FIR-radio correlation is
  comparable to the dispersion for $q_{70}$ on sub-kpc scales within
  the galaxy disks.  
  Also, the trend of increasing $q_{70}$ ratio with increasing
  70~$\mu$m luminosity within each galaxy is not observed in the {\it
  global} correlation.
\item
  The phenomenological modeling of cosmic ray electron (CR$e^{-}$)
  diffusion using an image-smearing technique is successful as it
  both decreases the measured dispersion in $q_{70}$ 
   and reduces the slope of the observed non-linearity in $q_{70}$
  with 70~$\mu$m surface brightness. 
   The latter suggests that the non-linearity may be due to the
  diffusion of CR$e^{-}$s from star-forming regions.  
\item
  Exponential kernels work marginally better to tighten the
  correlation than Gaussian kernels, independent of projection.
  This result suggests that CR$e^{-}$ evolution is not well described
  by a pure random-walk diffusion and requires additional processes
  such as escape and decay.
\item
  Our two less active star-forming galaxies require kernels having
  larger scale-lengths to improve the correlation while our two
  more active galaxies require smaller scale-lengths.
  This difference may be due to time-scale effects in which there has
  been a deficit of recent CR$e^{-}$ injection into the ISM of the two
  less active star-forming galaxies, thus leaving the underlying
  diffuse disk as the dominant structure in the morphology.
\end{enumerate}






%


\bibliographystyle{kapalike}

\begin{chapthebibliography}{<widest bib entry>}
\bibitem[Bicay \& Helou (1990)]{bh90} Bicay M. D. and Helou, G. 1990,
  ApJ, 362, 59
\bibitem[Dale \& Helou (2002)]{dd02} Dale, D. A. and Helou, G. 2002,
  ApJ, 576, 159
\bibitem[de Jong et al. (1985)]{de85} de Jong, T., Klein, U.,
  Wielebinski, R., and Wunderlich, E.  1985, A\&A, 147, L6
\bibitem[Harwit \& Pacini (1975)]{har75} Harwit, H, and Pacini, F
  1975, ApJL, 200, L127
\bibitem[Helou, Soifer, \& Rowan-Robinson (1985)]{gxh85} Helou, G.,
  Soifer, B. T., and Rowan-Robinson, M. 1985, ApJL, 298, L7
\bibitem[Kennicutt et al. (2003)]{rk03} Kennicutt, R. C. Jr., et al. 2003,
  PASP, 115, 928
\bibitem[Marsh \& Helou (1998)]{mh98} Marsh, K. A. and Helou, G. 1998,
  ApJ, 493, 121
\bibitem[Murphy et al. (2005)]{ejm05} Murphy, E. J., et al. 2005,
  ApJ, submitted
\bibitem[van der Kruit (1971)]{vk71} van der Kruit, P. 1971, A\&A, 15,
  110  
\bibitem[van der Kruit (1973)]{vk73} van der Kruit, P. 1973, A\&A, 29,
  263 
\bibitem[Yun, Reddy, \& Condon (2001)]{yun01} Yun, M. S., Reddy,
  N. A., and Condon, J. J.  2001, ApJ, 554, 803

\end{chapthebibliography}

\end{document}